\documentclass[aps,twocolumn]{revtex4}

\usepackage{amsmath}
\usepackage{amssymb}

\usepackage{graphicx}

\usepackage{color} 

\makeatletter
\renewcommand{\@biblabel}[1]{\quad#1.}
\makeatother

\newcommand{\ctst}{$\mathrm{C}_2\mathrm{S}_2$}
\newcommand{\ctstt}{$(\mathrm{C}_2\mathrm{S}_2)_2$}
\newcommand{\ctstm}{$\mathrm{C}_2\mathrm{S}_2\mathrm{M}$}
\newcommand{\ctstmt}{$\mathrm{C}_2\mathrm{S}_2\mathrm{M}_2$}
\newcommand{\ctstmx}{$\mathrm{C}_2\mathrm{S}_2\mathrm{M}_x$}

\begin{document}

\title{Coexistence between fluid and crystalline phases of proteins in photosynthetic membranes}
\author{Anna R.~Schneider}
\affiliation{Biophysics Graduate Group, University of California, Berkeley, CA, USA}
\author{Phillip L.~Geissler}
\affiliation{Department of Chemistry, University of California, Berkeley, CA, USA}
\affiliation{Chemical Sciences and Physical Biosciences Divisions, Lawrence Berkeley National Lab, Berkeley, CA, USA}
\email{geissler@berkeley.edu}

\begin{abstract}
Photosystem II (PSII) and its associated light-harvesting complex II (LHCII) are highly concentrated in the stacked grana regions of photosynthetic thylakoid membranes. Within the membrane, PSII--LHCII supercomplexes can be arranged in disordered packings, ordered arrays, or mixtures thereof. The physical driving forces underlying array formation are unknown, complicating attempts to determine a possible functional role for arrays in regulating light harvesting or energy conversion efficiency. Here we introduce a coarse-grained model of protein interactions in coupled photosynthetic membranes, focusing on just two particle types that feature simple shapes and potential energies motivated by structural studies. Reporting on computer simulations of the model's equilibrium fluctuations, we demonstrate its success in reproducing diverse structural features observed in experiments, including extended PSII--LHCII arrays. Free energy calculations reveal that the appearance of arrays marks a phase transition from the disordered fluid state to a system-spanning crystal, which can easily be arrested by thermodynamic constraints or slow dynamics. The region of fluid-crystal coexistence is broad, encompassing much of the physiologically relevant parameter regime. Our results suggest that grana membranes lie at or near phase coexistence, conferring significant structural and functional flexibility to this densely packed membrane protein system.
\end{abstract}

\maketitle


\section{Introduction}
Photosynthetic efficiency relies on precise spatial organization of 
sites for light harvesting, exciton transport, and charge separation \cite{Scholes2011}. In higher plants and green algae, these functions are carried out by pigment-proteins in the thylakoid membrane, with PSII hosting charge separation and oxygen evolution functionality and LHCII acting as its associated light-harvesting antenna. PSII and LHCII exhibit self-assembled organization on a range of length scales \cite{Dekker2005,Nevo2012}. At the macromolecular scale, dimers of PSII core complexes associate specifically with 1--6 trimers of LHCII and up to 2 monomers of each of the minor light harvesting complexes (CP24, CP26, and CP29) to form a family of PSII supercomplexes \cite{Boekema1999,Caffarri2009,Tokutsu2012}. On larger scales, the thylakoid membrane is differentiated into stacked discs of tightly appressed membrane called grana, tubes or sheets of unappressed membrane called stroma lamellae, and controversial junctional regions called margins \cite{Austin2011,Daum2011,Garab2008,Nevo2012}. PSII and LHCII are among the many proteins that display dramatic lateral heterogeneity within this complex membrane architecture---PSII and LHCII typically localize to appressed grana membranes, photosystem I (with its light-harvesting complex I) and ATP synthase are predominantly found in unappressed membranes, and other proteins such as cytochrome $b_6f$ are present in both regions \cite{Albertsson2001,Staehelin1995}. The degrees of both supercomplex formation and lateral heterogeneity are dynamically regulated in reponse to environmental factors by the processes of state transitions and PSII repair, in which phosphorylation of LHCII or PSII (respectively) is correlated with partial or complete dissolution of the supercomplex and migration of the phosphorylated protein species toward stroma lamellae or margin regions; many mechanistic details remain under debate \cite{Tikkanen2012,Vener2007,Minagawa2011,Dietzel2011}.

Within grana stacks, the most striking features of protein organization are regular 2D patterns termed arrays. Since their early description in the 1960s \cite{Park1964}, these motifs have frequently beeen observed in electron microscopy and atomic force microscopy studies of stacked thylakoid membranes, and are composed of tens or hundreds of unit cells containing one PSII core dimer and variable quantities of LHCII (reviewed in \cite{Kouril2012,Dekker2005,Hankamer1997}). Array extent and unit cell have been correlated with a number of experimental factors, including mutations of light-harvesting complex \cite{Yakushevska2003,Ruban2003,Kovacs2006,deBianchi2008,Damkjaer2009}, PsbS \cite{Kereiche2010}, or photosystem I \cite{Morosinotto2006} proteins; cold storage \cite{Semenova1995,Sznee2011}; and acclimation to low light \cite{Kirchhoff2007,Kouril2013}. Yet, in part because observed arrays can vary widely between similar samples \cite{Boekema2000,Yakushevska2001}, quantitatively predictive statements about the structure and function of PSII arrays have not emerged. Hypothesized functional rationales for PSII arrays primarily focus on predicted exciton transport effects, such as optimizing antenna size, excitation quenching site availability, or reaction center connectivity to match light conditions \cite{Croce2011,Horton2008}, and on protein and electron carrier mobility arguments \cite{Kirchhoff2008,Goral2012}, but no functional role has been proven.

Nanoscale maps of the relative arrangement of LHCII and PSII in native-like grana membrane environments will be necessary for building a complete picture of exciton flow during photosynthetic light harvesting and energy conversion. The oxygen-evolving complex of PSII protrudes above the plane of the membrane on the lumenal side \cite{Zouni2001,Kamiya2003,Ferreira2004,Umena2011,Kern2012}, allowing it to be easily identified in electron microscopy and atomic force microscopy images of grana membranes. LHCII lacks a lumenal protrusion \cite{Liu2004,Standfuss2005}; this has prevented microscopists from simultaneously mapping the positions of LHCII and PSII in disordered membrane environments (i.e., outside of arrays). Super-resolution fluorescence imaging techniques are a promising alternative, but have not yet been successfully applied to grana membranes, which are already highly fluorescent. Computer simulation has the potential to complement experiment and address this important gap in structural and mechanistic understanding.

The first computational study of thylakoid protein organization focused on a single membrane layer, modeled as a flat 2D sheet, and treated grana and stroma lamellae membrane proteins as hard disc-shaped particles moving within that sheet \cite{Drepper1993}. Tremmel and coworkers developed a distinct model restricted to grana membrane protein complexes (LHCII trimers, PSII supercomplexes, and cytochrome $b_6f$) that used single-particle Monte Carlo moves to sample configurations of hard particles with highly detailed shapes on a single 2D lattice \cite{Tremmel2003,Kirchhoff2004}; they later extended the model by introducing short-range patchy interactions between protein particles \cite{Tremmel2005}, a standard means of modeling anisotropic protein-protein interactions \cite{Sear1999,Kern2003,Whitelam2009,Bianchi2011,Haxton2012}. None of these publications reported the emergence of PSII arrays in computer simulation.

In this study, we use a simple nanoscale computational model of LHCII and PSII in stacked membranes to analyze grana protein organization. In particular, we examined PSII array formation in system sizes comparable to full grana discs (hundreds of nanometers in diameter). We extend previous computational approaches in two key ways: we simulate multiple, coupled membrane layers in a grana stack; and we employ computational methods capable of exploring highly cooperative transitions. By thoroughly examining thermal fluctuations in our model of a PSII--LHCII binary mixture, we reveal a phase transition between a relatively dilute, disordered PSII--LHCII fluid and a dense, ordered PSII--LHCII crystal. Physiological protein concentrations are at fluid-crystal phase coexistence or near the boundary between the fluid and coexistence regions, where small changes in protein density or interactions can lead to dramatic shifts in the observed degree of array formation and in any array-dependent functionality.

\section{Results}

\subsection{Model is founded on in vivo phenomenology}

\begin{figure}
\begin{center}
\includegraphics[width=8cm]{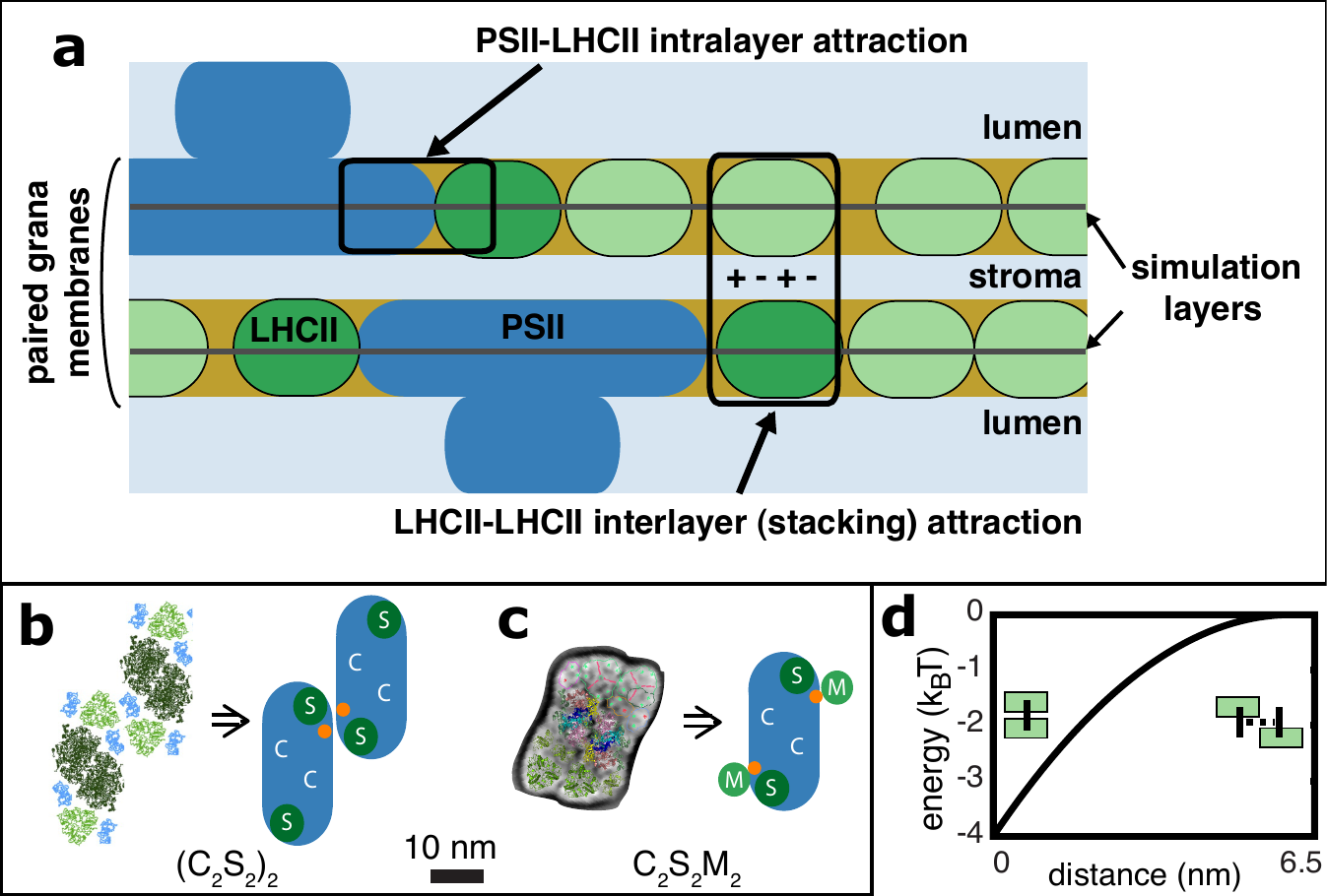}
\end{center}
\caption{
{\bf Graphical summary of the model.}
\textbf{a}: Side view cartoon of protein geometry and interactions. PSII (blue) and LHCII (green) particles reside in two coupled layers, each of which represents a lipid bilayer grana membrane (tan) surrounded by stromal and lumenal aqueous regions (pale blue).
\textbf{b} and \textbf{c}: Top view from the lumenal side of the coarse-graining procedure. Solved protein structures of LHCII trimers are roughly circular; model LHCII particles are hard discs with diameter 6.5 nm (dark green: S-LHCII; medium green: M-LHCII). PSII \ctst~supercomplexes are roughly rod-shaped; model PSII particles are hard discorectangles with diameter 12 nm and tip-to-tip length 26.5 nm (blue). The orange dots show LHCII binding sites that allow formation of supramolecular complexes, including \ctstt~(\textbf{b}, governed by $\epsilon_{\mathrm{SL-P}}$) and \ctstmt~(\textbf{c}, governed by $\epsilon_{\mathrm{ML-P}}$). Protein structures adapted from \cite{Caffarri2009, Dekker2005}. 
\textbf{d}: Potential energy of the LHCII interlayer stacking interaction. Insets: side view cartoons of unbound LHCII (light green) in stacked (left) and unstacked (right) configurations.}
\label{fig:model}
\end{figure}

Model details are illustrated in Fig.~\ref{fig:model} and Fig.~S1, and further described in the Methods and Text S1. Our model consists of two coupled 2D layers, representing stroma-paired grana membranes, which comprise the minimal system for PSII array formation \cite{Daum2010}. These two layers can be thought of as an isolated membrane pair, or as two stroma-paired membranes in a larger grana stack where lumen-side interactions are assumed to be negligible \cite{Dekker2005}. Two types of hard particles reside in these layers: disc-shaped species that represent LHCII trimers, and rod-shaped species that represent PSII supercomplexes (specifically the so-called \ctst~supercomplex in \emph{Chlamydomonas} \cite{Tokutsu2012}, spinach \cite{Nield2000}, and \emph{Arabidopsis} \cite{Caffarri2009}). PSII particles, like PSII protein supercomplexes, have two constituent embedded LHCIIs that move with the PSII particle as a rigid body (Fig.~\ref{fig:model}b and c, Methods). The embedded LHCIIs are constrained to their physiological locations on either side of the PSII axis \cite{Nield2006}, making the PSII particles chiral (Fig.~S1).

In addition to excluding area within the layer that they occupy, these particles attract one another in three distinct ways. One mode of attraction acts between LHCII particles in opposing layers. Electrostatic interactions between LHCII stromal faces contribute to grana stack integrity \cite{Barber1982} and are implicated in PSII array formation in experiments \cite{Daum2010,Boekema2000}. Standfuss and coworkers have proposed a ``velcro-like'' qualitative model for the LHCII stacking interaction \cite{Standfuss2005}, but a quantitative understanding is lacking. We have chosen a simple phenomenological form for the LHCII--LHCII interlayer attraction that favors LHCII face-to-face contact (Fig.~\ref{fig:model}a,d). We set the strength of the attraction to $\epsilon_{\mathrm{L-L}} = 4$ $k_{\textrm{B}}T$, which creates strong but reversible binding and is consistent with electrostatic measurements (see Text S1). All LHCII particles can participate in this interaction.

The other attractions associate PSII and LHCII particles residing in the same layer. When protein complexes are isolated from thylakoid membranes of \emph{Arabidopsis}, spinach, or \emph{Chlamydomonas}, LHCII complexes are found to be distributed among at least three different local environments: they can be strongly bound within a supercomplex (embedded or S-LHCII), moderately bound to the edge of a supercomplex (bound or M-LHCII), or structurally detached from PSII (free LHCII) \cite{Caffarri2009,Dekker2005,Yakushevska2001,Tokutsu2012}. Motivated by the apparent equilibrium between moderately bound and free LHCIIs, we introduce two interaction sites on the periphery of each PSII particle, as illustrated by the orange dots in Fig.~\ref{fig:model}b and c and Fig.~S1. These interaction sites attract LHCIIs in the same layer over a short range ($\approx$1 nm), such that modeled M-LHCIIs in the ``bound'' state are constrained to locations relative to PSII that are consistent with published \ctstmx~complexes \cite{Caffarri2009,Tokutsu2012}.

Experiments suggest that the intralayer interactions between PSII supercomplexes and free LHCIIs are similar to yet distinct from those between a PSII supercomplex and the S-LHCII of another supercomplex; for instance, binding of M-LHCII (Fig.~\ref{fig:model}c) appears to require PSII subunit CP24 in \emph{Arabidopsis}, while association of S-LHCII with the same site (Fig.~\ref{fig:model}b) may not \cite{Kovacs2006}. We therefore define separate energy scales for these interactions ($\epsilon_{\mathrm{ML-P}}$ and $\epsilon_{\mathrm{SL-P}}$, respectively). Based on measurements of intramembrane associations between other protein complexes in photosynthetic membranes \cite{Liu2011}, these energies are expected to be modest, on the order of one or a few $k_{\mathrm{B}}T$; we focus on $\epsilon_{\mathrm{SL-P}} = \epsilon_{\mathrm{ML-P}} = 2 k_{\mathrm{B}}T$.

We emphasize that this model includes only the short-ranged intramembrane protein-protein attractions that are most conclusively established by single-particle structural studies. Longer-range patterns in our simulations can only arise from emergent correlations involving many particles. We also emphasize that the values of the interaction strengths that we have selected are intended to be representative of the physiology of a generic grana membrane, and not to represent any specific species or growth condition. Experimental measurements of the interaction strengths, which would be difficult but not inconceivable (e.g., along the lines of Ref.~\cite{Liu2011}), could support future efforts to tailor the model to precisely match specific experimental conditions.

\subsection{Simulations capture experimentally-determined intramembrane protein organization}
We performed Monte Carlo simulations to sample equilibrium particle configurations at a wide range of particle concentrations within experimentally relevant parameter regimes. Two standard experimental measures of grana protein content are the LHCII:PSII ratio (specifically, the mole ratio $\phi$ of free LHCII trimers to PSII \ctst~supercomplexes) and the number density of PSII (typically PSII per square micron of membrane). These metrics can be combined to estimate the total protein packing fraction $\rho$ (i.e., the fraction of grana membrane area occupied by proteins). Published values of these quantities vary significantly based on plant growth conditions and membrane preparation protocols: $\phi$ is typically in the range $\approx$ 2--6 \cite{Jansson1997,Kirchhoff2007,Veerman2007}, and $\rho$ is typically in the range $\approx$ 0.6--0.8 \cite{Kirchhoff2004,Haferkamp2010,Kouril2013}.

\begin{figure}[!ht]
\begin{center}
\includegraphics[width=8cm]{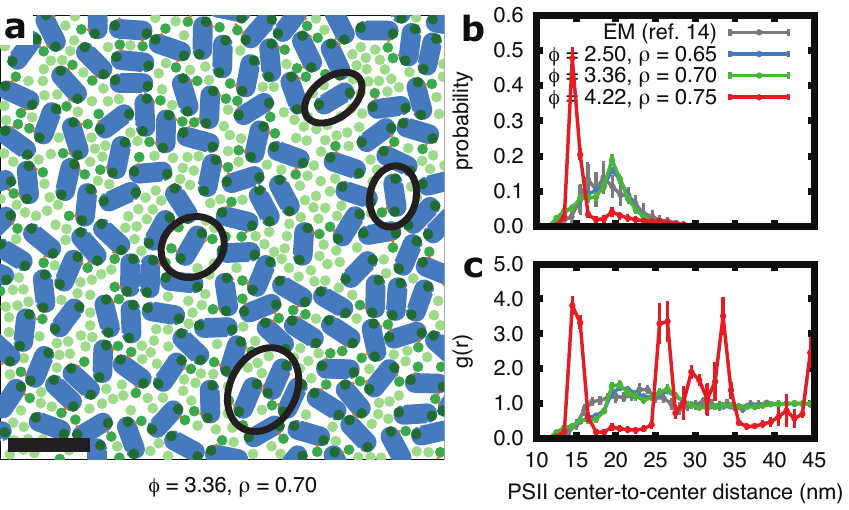}
\end{center}
\caption{
{\bf Intramembrane structure of a coupled pair of model grana membranes at 1750 PSII/$\mu m^{2}$.} \textbf{a}: Snapshot of the lumen-side-up membrane layer from a simulation at $\rho = 0.70$, $\phi = 3.36$ showing a representative disordered configuration. Example supramolecular complexes are circled, clockwise from top: \ctst, \ctstm, \ctstt, \ctstmt. Color scheme as in Fig.~\ref{fig:model}; scale bar = 50 nm. \textbf{b} and \textbf{c}: Comparison between experimental and simulated data for the statistics of intramembrane PSII center-to-center separation distances. \textbf{b} shows the distribution of nearest-neighbor distances, and \textbf{c} shows the radial distribution function $g(r)$. Freeze-fracture EM data on isolated spinach grana membranes \cite{Kirchhoff2004}, with 1700
PSII/$\mu m^{2}$ on average, are consistent with these simulations of the disordered state. In contrast, simulations at $\rho = 0.75$ show PSII aggregation and ordering.}
\label{fig:structure}
\end{figure}

Simulations of moderately dense systems ($\rho = 0.70$, $\phi = 3.36$) showed good agreement with a variety of experimental probes of intramembrane protein organization. Fig.~\ref{fig:structure} shows data from lumen-side-up layers of simulated membrane stacks, the perspective from which PSII positions are routinely detected experimentally \cite{Kouril2012}. LHCII and PSII particles were distributed uniformly and isotropically throughout the layer, with PSII orientations correlated over about 30 nm (2.5 PSII widths). Statistics of PSII--PSII separation distances matched well with experimental results, with computed and experimental nearest-neighbor PSII center-to-center distances of $18.7 \pm 2.9$ nm and $19.0 \pm 4.1$ nm, respectively, and weakly structured radial distribution functions $g(r)$ (Fig.~\ref{fig:structure}b and c, \cite{Kirchhoff2004}); small offsets in peak positions could be due to our simplified particle shapes. These correlations arise both from direct association between PSII pairs and from effective forces mediated by LHCIIs and by particles in the opposing layer. Contributions from LHCII fluctuations include ``depletion'' attractions \cite{Asakura1958}, which tend to cluster PSIIs in order to maximize space available for the smaller LHCII species to explore.

Though simple in form, our model of PSII--LHCII intralayer attraction was sufficient to stabilize significant populations of several known supramolecular complexes---PSII supercomplex dimers (\ctstt), PSII supercomplex with one additional LHCII (\ctstm), and PSII supercomplex with two additional LHCIIs (\ctstmt)---in equilibrium with a pool of lone PSII supercomplexes (\ctst). The balance of this equilibrium can be tuned by changing the strength of the intralayer attraction; at $\epsilon_{\mathrm{SL-P}} = \epsilon_{\mathrm{ML-P}} = 2$ $k_{\textrm{B}}T$, LHCII binding and unbinding is facile, and approximately 25\% of LHCII particles are bound to PSII supercomplexes under these conditions.

\subsection{LHCII stacking creates interlayer PSII correlations}
\begin{figure}
\begin{center}
\includegraphics[width=8cm]{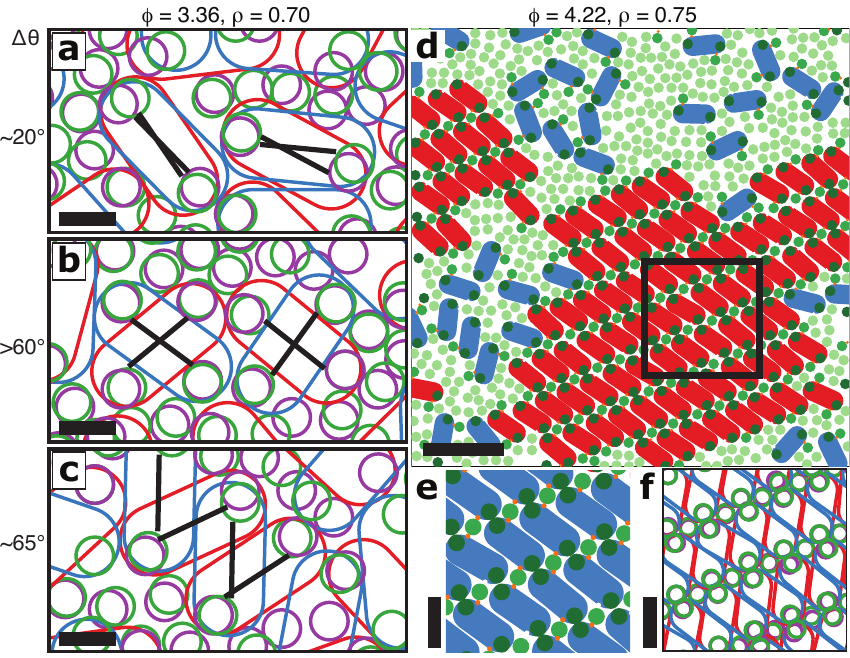}
\end{center}
\caption{
{\bf LHCII stacking effects on PSII organization.}
\textbf{a}-\textbf{c}: PSII interlayer motifs, as described in the text. Green and blue outlines: LHCII and PSII particles in the upper (lumen-side-up) layer; purple and red outlines: LHCII and PSII particles in the lower (stroma-side-up) layer. Black lines drawn parallel to the long axis of selected rods highlight orientational relationships between PSII particles in different layers. Orientation correlations $\Delta \theta$ also appear in Fig.~S2c. Scale bars = 10 nm.
\textbf{d}: Snapshot of the top layer from a simulation at $\phi = 4.22$, $\rho = 0.75$ showing a representative array. Color scheme as in Fig.~\ref{fig:model}, except arrayed PSII are colored red (see Methods). Scale bar = 50 nm.
\textbf{e} and \textbf{f}: Magnified views of the boxed region in panel \textbf{d}. Scale bars = 20 nm. The role of intralayer attractions is highlighted in \textbf{e} by showing only particles in the top layer and indicating the locations of PSII--LHCII interaction sites on each PSII. The stabilizing role of stacking is highlighted in \textbf{f} by showing particles from both layers in outline form (as in panels \textbf{a}-\textbf{c}).
}
\label{fig:arrays}
\end{figure}

Our simple model of LHCII stacking interactions similarly proved sufficient to generate a range of structural motifs that have been inferred from experiment. Specifically, we frequently observed (1) pairs of stacked, nearly parallel PSIIs, in which the LHCIIs chirally embedded in one PSII are both aligned with the embedded LHCIIs of the opposing PSII (of opposite chirality) (Fig.~\ref{fig:arrays}a); (2) nearly perpendicular pairs, in which all embedded LHCIIs stack over free LHCIIs in the other layer (Fig.~\ref{fig:arrays}b); and (3) rows of rotated PSIIs in one layer stacked atop a parallel but oppositely rotated row in the other layer, in which the two LHCIIs embedded in a given PSII are aligned with those of two different PSIIs in the opposing layer (Fig.~\ref{fig:arrays}c). In the absence of LHCII stacking interactions (i.e., when $\epsilon_{\mathrm{L-L}}=0$), these PSII correlations disappeared (Fig.~S2).

All three of these motifs have been reported in separate experimental studies \cite{Boekema2000,Daum2010,Nield2000}. From those observations, however, it was not clear whether different motifs could be simultaneously abundant; nor could those authors establish LHCII stacking as a sufficient driving force for various interlayer correlations. Our results indicate that LHCII stacking can indeed drive all of these associations among PSIIs in opposing layers. 

\subsection{Simulated PSII arrays depend on packing fraction and attraction strength}
Above a packing fraction of $\rho \approx 0.7$, simulations exhibited sizable ordered arrays, featuring alternating rows of PSIIs and LHCIIs (Fig.~\ref{fig:arrays}d-f). Interestingly, PSIIs in these configurations do not engage in direct intralayer attractions. Instead, adhesion between PSII rows is provided by intralayer attractions to the interspersed M-LHCIIs that bridge between PSII rows. Each row of PSIIs is stabilized by stacking interactions with a PSII row in the opposing layer, as in Fig.~\ref{fig:arrays}c (and by the less geometrically specific depletion attractions). This key role for stacking of embedded LHCIIs is highlighted by an absence of ordered arrays in simulations with $\epsilon_{\mathrm{L-L}}=0$ (Fig.~S3). The experimental correlation between arrays and stacking \cite{Daum2010} supports the realism of our coarse-grained model, as well as the conclusion that both modes of attraction in our model are essential for ordering.

\subsection{A fluid-crystal phase transition is manifested in osmotic ensemble simulations}
Configurations in which tens or hundreds of PSIIs cluster tightly together, such as in Fig.~\ref{fig:arrays} and in many EM images, reflect strong emergent forces of association, sufficient to offset the entropic cost of sequestering the constituent complexes from the surrounding disordered environment. Computer simulations allow us to address whether these large arrays further signify a more profound underlying phenomenon, namely, a true phase transition from a disordered ``fluid'' of relatively low PSII density to a system-spanning crystal of tightly packed PSIIs. The simulations we have described thus far are not suited to address this question, nor would be experiments probing isolated grana membranes with fixed protein content. In these situations crystallization could not proceed to completion, simply because the system's net composition and total area are fixed at values inconsistent with the crystalline phase. Indeed, in simulations of closed systems like those shown in Fig.~\ref{fig:arrays}, array growth halted before all PSII had been incorporated, creating a dynamic equilibrium between arrayed and disordered PSII (Fig.~S4).

Phase transitions can be more readily identified by studying open systems such as the osmotic ensemble, in which area and composition can fluctuate subject to external fields at fixed temperature \cite{Mehta1994}. First, we imposed a 2D ``pressure'' $p$ that regulates changes in total area. More precisely, we fixed the osmotic pressure that particles experience parallel to the plane of the membrane; simulated area fluctuations implicitly represent addition or subtraction of lipids from the membrane. Second, we allowed the population of one component (we chose LHCII) to fluctuate at fixed chemical potential $\mu_{\rm L}$. Corresponding number fluctuations in a real system involve exchanging material with a very large bath. In intact thylakoids, the stroma lamellae and other connected grana stacks could play the role of a bath.

\begin{figure*}
\begin{center}
\includegraphics[width=15cm]{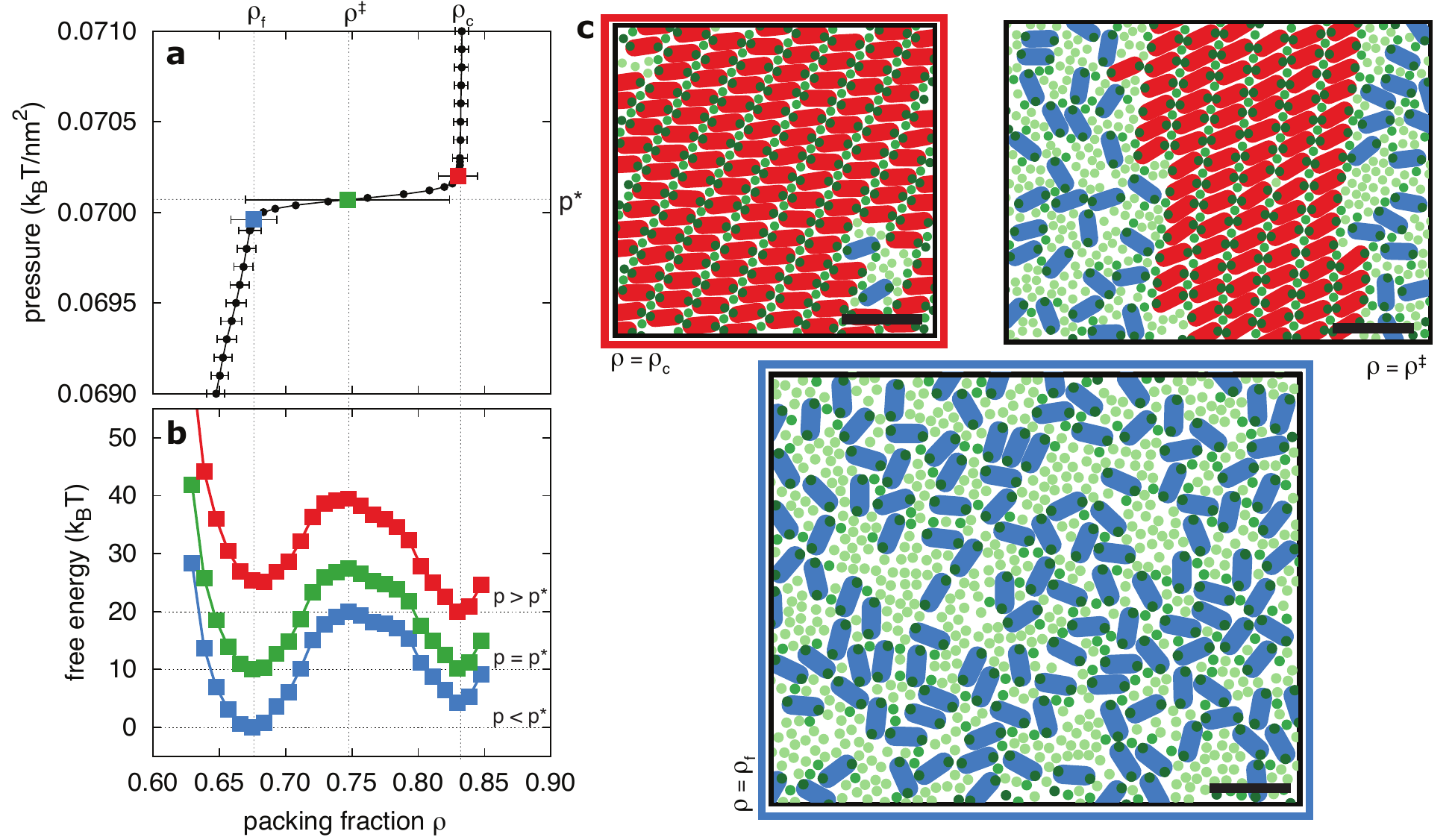}
\end{center}
\caption{
{\bf Umbrella sampling simulations provide evidence for a phase transition.} \textbf{a}: Applied pressure $p$ versus packing fraction $\rho$ along a line of constant chemical potential shows a sharp crossover at $p^{*}$ from a low-pressure, low-packing fraction regime to a high-pressure, high-packing fraction regime. Means (line and points) and root-mean-squared fluctuations (whiskers) of $\rho$ at each pressure are calculated from probability distributions derived from free energy surfaces like those in panel \textbf{b}. Fluctuations, and therefore whiskers, are large in the vicinity of $p^{*}$; for clarity we show only one whisker in this region.
\textbf{b}: Free energy as a function of $\rho$ for a system at relative chemical potential $\bar{\mu}_{\textrm{L}} = 0.1$ $k_{\textrm{B}}T$ and three values of pressure near coexistence: within the fluid phase (blue, $p=0.06996$ $k_{\textrm{B}}T/\textrm{nm}^{2}$), within the crystalline phase
(red, $p=0.07020$ $k_{\textrm{B}}T/\textrm{nm}^{2}$), and at coexistence (green, $p^{*}=0.07007$ $k_{\textrm{B}}T/\textrm{nm}^{2}$). Error bars estimated from the MBAR method are smaller than the symbols. Because the zero of free energy is arbitrary at each pressure, curves are vertically offset for clarity. Dotted lines are guides to the eye.
\textbf{c}: Snapshots taken from umbrella sampling simulations biased to the stated packing fractions. Color scheme as in Fig.~\ref{fig:arrays}d. Scale bars = 50 nm.}
\label{fig:c2s2m_phasetrans}
\end{figure*}

Our examination of detailed phase behavior focused on the type of array shown in Fig.~\ref{fig:arrays}d by restricting the model interactions. In the simulations described below, intralayer attractions between PSIIs and embedded LHCIIs were omitted by setting $\epsilon_{\mathrm{SL-P}} = 0$, while attractions to free LHCIIs are retained by maintaining $\epsilon_{\mathrm{ML-P}} = 2$ $k_{\textrm{B}}T$; thus, \ctstmx~complexes were the only single-layer supramolecular structures directly stabilized by model energetics. Since the intralayer attractions we disabled do not directly contribute to the stability of these arrays, we expect the fully interacting model to exhibit qualitatively similar phase behavior. The prevalence of \ctstmx~complexes in \textit{Arabidopsis} \cite{Caffarri2009} and \textit{Chlamydomonas} \cite{Tokutsu2012} suggests that the restricted model may be particularly appropriate for these organisms.

Varying pressure at fixed chemical potential $\mu_{\textrm{L}}$ produced a sharp change in average packing fraction (Fig.~\ref{fig:c2s2m_phasetrans}a), indicating a highly cooperative transition. The concomitantly sudden appearance of a system-spanning PSII array (Fig.~\ref{fig:c2s2m_phasetrans}c, Fig.~S6) suggests that the degree of cooperativity would grow with system size, as in a first-order phase transition, and identifies the high-pressure phase as crystalline. For the finite, micron-scale system that we simulated, the jump in packing fraction $\rho$ is necessarily rounded and could only become discontinuous in the thermodynamic limit of an infinitely large system. Demonstrating true phase behavior would require an analysis of scaling as this limit is approached. Given the limited spatial extent of natural thylakoids, we instead scrutinized remnant hallmarks of phase coexistence in a finite system, specifically bistable free energy profiles and the presence of stable interfaces.

We computed the free energy $F(\rho)$ as a function of packing fraction for specific values of the thermodynamic parameters ($p$, $\mu_{\textrm{L}}$, $T$) using umbrella sampling (Fig.~S5). Free energy profiles at many other values of these parameters were then calculated by thermodynamic reweighting \cite{Shirts2008}. These profiles exhibit two distinct basins, with minima at $\rho_{\rm f} < 0.7$ and $\rho_{\rm c} > 0.8$, over a range of pressures (Fig.~\ref{fig:c2s2m_phasetrans}b). The low-packing fraction minimum corresponds to a disordered PSII--LHCII two-component fluid (bottom of Fig.~\ref{fig:c2s2m_phasetrans}c); configurations representative of the high-packing fraction minimum are nearly perfect PSII--LHCII co-crystals (top left of Fig.~\ref{fig:c2s2m_phasetrans}, Fig.~S6). Constraining the packing fraction to lie midway between $\rho_{\rm f}$ and $\rho_{\rm c}$ produces heterogeneous structures in which fluid and crystal coexist, separated by a system-spanning interface (Fig.~\ref{fig:c2s2m_phasetrans}c). Together with observations of hysteresis when pressure is cycled above and below the coexistence pressure $p^*$ (Fig.~S7), these observations point strongly to a first-order phase transition.

\subsection{Phase coexistence region includes physiological conditions}
\begin{figure}[!ht]
\begin{center}
\includegraphics[width=8cm]{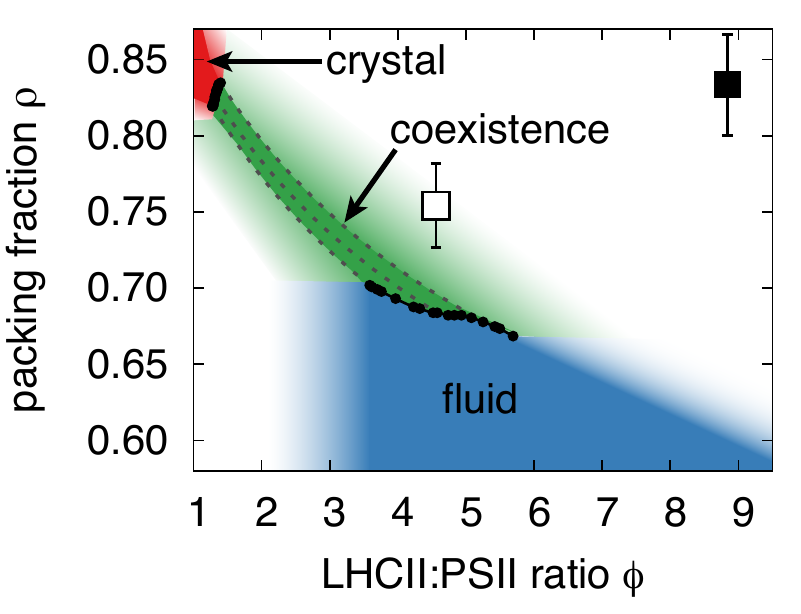}
\end{center}
\caption{
{\bf Phase diagram of crystal--fluid coexistence in the ($\phi$, $\rho$) plane.} Each thermodynamic state ($\mu_{\textrm{L}}^{*}$, $p^{*}$) of phase equilibrium maps in this plane to two points and a corresponding tie line.  One point lies at the boundary of the homogeneous fluid phase (low packing fraction and high LHCII content, shaded blue), the other at the boundary of the fully crystalline state (high packing fraction and low LHCII content, shaded red). Example tie lines are drawn as dashed lines connecting these boundaries for selected values of $p^{*}$. Coexistence (shaded green) and pure-phase regions extend beyond the shaded areas determined by our limited data. In particular, the coexistence region is expected to span the entire upper-right-hand quadrant. Filled and open symbols are low-light and ordinary-light phase points, respectively, calculated from data in Ref.~\cite{Kirchhoff2007} (see SI Methods).}
\label{fig:phasediag}
\end{figure}

From free energy profiles like those of Fig.~\ref{fig:c2s2m_phasetrans}b, we constructed a phase diagram in the plane of packing fraction $\rho$ and composition $\phi$. For a given chemical potential $\mu_{\textrm{L}}^{*}$, the coexistence pressure $p^{*}$ can be uniquely identified as the pressure that maximizes the variance of packing fraction fluctuations; note the large standard deviation at $p^{*}$ in Fig.~\ref{fig:c2s2m_phasetrans}a. Values of $\rho$ and $\phi$ for the two phases at the thermodynamic state ($\mu_{\textrm{L}}^{*}$, $p^{*}$) determine a pair of points at the boundaries of the crystal--fluid coexistence region. Repeating this procedure for different values of $\mu_{\textrm{L}}^{*}$, we traced out coexistence curves bounding regions where homogeneous fluid and crystal phases are thermodynamically stable (Fig.~\ref{fig:phasediag}).

The dashed lines in Fig.~\ref{fig:phasediag} trace tie lines, i.e., lines along which the relative extent of the two pure phase regions varies while the packing fraction and composition of each phase remains constant (as determined by the endpoints). These tie lines thus enable straightforward predictions for thermodynamic states in the coexistence region, requiring no characterization beyond the physical properties of the endpoints. 

Remarkably, the resulting coexistence region encompasses the grana packing fractions and compositions reported from many in vivo and in vitro experiments (e.g., filled and open squares in Fig.~\ref{fig:phasediag}). Other experiments report values of $(\rho, \phi)$ within the model's fluid phases, but not far from the coexistence curve. Thus, our model of the LHCII--PSII protein system supports coexistence in many physiologically relevant conditions.

\section{Discussion}

We have demonstrated that simulations of a simple model of PSII and LHCII in stacked grana membranes, when configured to represent a generic grana membrane, can recapitulate many disparate and nontrivial experimental observations. These behaviors emerge spontaneously from the model's short-ranged interactions without the need for us to presuppose any particular target assembled structures. Specifically, we observe a distribution of \ctst, \ctstm, and \ctstmt\ complexes (Fig.~\ref{fig:structure}a); LHCII-mediated intermembrane associations between PSII supercomplexes (Fig.~\ref{fig:arrays}a--c); and co-occurence of disordered and crystalline-ordered regions in PSII- and LHCII-rich membranes (Fig.~\ref{fig:arrays}d). Importantly, all of these qualitative features appear to be common to grana membranes from many photosynthetic organisms and many growth conditions, although the quantitative details of course vary. To the best of our knowledge, this work marks the first reported computational investigation of grana membranes to share these commonalities with experiment.

Our principal prediction from numerical studies of this model is that the appearance of finite arrays of PSII and LHCII signals thermodynamic coexistence of disordered (fluid) and ordered (crystalline) phases. The phase boundaries we have computed further suggest that many physiological conditions lie at or near such coexistence. The experimental data reported in Ref.~\cite{Kirchhoff2007}, correlating degree of array formation with the packing fraction and composition of grana membranes, offer the most concrete opportunity for comparison. Membranes at conditions just inside our coexistence region, from plants grown in ``ordinary'' light, were found to exhibit a low degree ($<2 \%$) of crystallinity. Membranes corresponding to conditions deep within our coexistence region, grown in low light, showed significantly enhanced ($22\%$) crystallinity. These consistent observations are indicated by the filled and open symbols in Fig.~\ref{fig:phasediag}. The tie line we have computed suggests that a fully equilibrated system in ordinary light would possess greater crystallinity than observed in experiment, possibly reflecting high nucleation barriers and/or slow growth characteristic of dynamics near coexistence (see Fig.~S6). A recent study of low-light-acclimated membranes \cite{Kouril2013} reported arrays at protein packing fractions ($\rho<0.6$) much lower than previously measured. Our calculations do not provide a direct way to resolve this apparent discrepancy, unless the different organisms feature substantially different energy scales and/or modes of protein association.

Experiments that systematically quantify dependence of crystallinity upon packing fraction at various protein compositions could further test or exploit our predictions. For example, grana membranes could be isolated from plants grown at different light intensities, generating samples over a range of values of $\phi$. Diluting these membranes with additional lipid [54] and measuring crystallinity via AFM or EM would allow construction of experimental phase diagrams analogous to those we have computed. Matching experimental and theoretical results in detail could determine appropriate values of the interaction parameters in our model.

The well-understood phenomenology of phase transitions helps explain why the extent of PSII array formation varies so dramatically between similar samples in experiment. The presence, extent, and number of arrays in the laboratory may be influenced by the characteristically slow dynamics associated with phase transitions. Nucleation of one phase from the other will be governed by rare structural fluctuations; crowding within grana membranes \cite{Kirchhoff2008,Kirchhoff2011} will likely make the subsequent repartitioning of material between phases slow as well. Moreover, grana isolation protocols that increase the protein packing fraction beyond the freezing transition densities for 2D hard discs ($\rho \approx 0.7$ \cite{Huerta2006}) or hard rods ($\rho \approx 0.8$ \cite{Bates2000}), such as BBY \cite{Haferkamp2010}, could trap the isolated grana in jammed nonequilibrium configurations, further complicating experimental determinations of the equilibrium distribution of PSII arrays. Further computational work will be required to elucidate these dynamic effects.

Proximity to phase coexistence could also contribute to substantial changes in thylakoid function observed to accompany modest changes in protein content and interactions. In addition to the low-light acclimation scenario discussed above, many regulatory processes associated with non-photochemical quenching, photoprotection, and repair shift the system's position relative to phase boundaries in the $\rho,\phi$-plane: (1) State transitions involve transport of LHCII from PSII-rich to PSI-rich membranes \cite{Iwai2010}, reducing the local protein density and LHCII:PSII ratio. (2) Photoinhibition-induced phosphorylation decreases the diameter of grana stacks and breaks up PSII supercomplexes \cite{Fristedt2009,Herbstova2012}, changing the system size and composition. (3) The qE component of nonphotochemical quenching may introduce an additional intramembrane attraction among LHCII \cite{Johnson2011}, affecting the relative stability of the fluid and crystal phases. These spatioregulatory processes are often interpreted as acting primarily over short length scales, tuning exciton fate by changing the relative distances between light harvesting sites, quenchers, and reaction centers. We suggest that, in addition, these processes may directly regulate global protein organization within the thylakoid membrane by inducing cooperative structural transitions.

Suitably adapted, the model and computational framework developed in this study may help to clarify the mechanisms of such spatioregulatory changes. Direct analyses of grana-scale LHCII organization are difficult and rare, with the notable recent exception of Ref.~\cite{Johnson2011}. With a few experimental parameters as input (namely $\phi$ and $\rho$ or their equivalents), our methods can generate physiologically reasonable LHCII configurations in the presence of PSII in a physically grounded and unbiased fashion, and in sufficient quantity to enable statistical comparisons (e.g., Figs.~S2 and S3). These ensembles of configurations could serve as the foundation for future studies of the many phenomena in which LHCII plays a key role.

\section{Model and Methods}

\subsection{Model geometry and energetics}
We represent protein complexes within membrane layers of a grana stack as greatly simplified particles that can move in only two directions ($x$ and $y$) and rotate only about the axis $z$ perpendicular to these layers. Our simulations sample configurations of a pair of these two-dimensional systems, periodically replicated in $x$ and $y$, coupled by stacking interactions between particles in different layers. The model includes two particle species: isotropic disc-shaped particles represent trimeric LHCII complexes, and  discorectangle-shaped particles represent PSII \ctst~supercomplexes. Particle shapes are simple approximations of the solved protein complex structures \cite{Caffarri2009,Nield2006}, and particle sizes are assigned to be consistent with the protein structures and with previous coarse-grained models \cite{Tremmel2003}. Specifically, each LHCII particle has a diameter $\sigma_{\textrm{L}}$ = 6.5 nm, and each PSII particle has a rectangle width and cap diameter $D_{\textrm{P}}$ = 12.0 nm and rectangle length $L_{\textrm{P}}$ = 14.5 nm. The locations of the LHCIIs embedded in the PSII particles are modeled on Ref.~\cite{Nield2006}. Full geometry details are given in Fig.~S1.

In addition to steric repulsions, our model includes three types of
attractive interactions. The first two attractions (with energy scales $\epsilon_{\textrm{SL-P}}$ and $\epsilon_{\textrm{ML-P}}$) act between an interaction site on PSII and an LHCII particle. The interaction is square-well, with a distance cutoff of $0.66 \sigma_{\textrm{L}}$ between the LHCII center and the binding site,  or equivalently a distance cutoff between the LHCII edge and the binding site of $0.16 \sigma_{\textrm{L}} = 1.04$ nm; this short-range distance cutoff ensures that each binding site can bind at most one LHCII at a time. This square well completely describes the $\epsilon_{\textrm{ML-P}}$ interaction between PSIIs and nonembedded LHCII particles. The attraction between PSIIs and embedded LHCIIs,  $\epsilon_{\textrm{SL-P}}$, has an additional constraint: the angle difference between the long axes of the two PSII particles hosting the interaction sites must be $\leq 30^{\circ}$. The depth of the square wells are set at $\epsilon_{\textrm{SL-P}} = \epsilon_{\textrm{ML-P}} = 2$ $k_{\textrm{B}}T$, except where noted. Temperature $T$ fixed at 1 throughout, such that $k_{\textrm{B}}T$ is our reduced unit of energy.

The other attraction involves two LHCII particles (free or embedded) in different layers. This interaction has an energetic minimum, with a value of $\epsilon_{\textrm{L-L}}$, when one disc completely eclipses another as viewed from above the layers. It vanishes continuously at $r=\sigma_{\textrm{L}}$, where $r$ is the lateral distance between the
centers of two LHCII discs. This dependence is plotted in Fig.~\ref{fig:model}d and its functional form is given in the SI
Methods. The energy scale is set to $\epsilon_{\textrm{L-L}} =
4$ $k_{\textrm{B}}T$, except where noted.

\subsection{Monte Carlo simulations}
Using standard Metropolis Monte Carlo methods, we sampled equilibrium configurational distributions of systems comprising $N_{\textrm{P}} = 2\times 128$ PSII particles and a variable number $N_{\textrm{L}}$ of free LHCII particles.  Trial moves included translational displacements of a randomly chosen LHCII or PSII particle, and rotational displacements of a randomly chosen PSII particle. Translational moves were restricted to periodically replicated 2D surfaces, each with area $A$ per layer. $x$- and $y$-components of translational moves were chosen from Gaussian distributions with zero mean and variances of 0.6 nm$^2$ for free LHCII displacements or 0.2 nm$^2$ for PSII displacements. PSII rotations were chosen uniformly on the interval $\left(-10^{\circ}, 10^{\circ}\right)$. An ``MC sweep'' consisted on average of $N_{\textrm{L}}$ LHCII displacements, $N_{\textrm{P}}$ PSII displacements, and $N_{\textrm{P}}$ PSII rotations.  We performed simulations of closed systems (i.e., sampling the canonical ensemble) with several values of $N_{\textrm{L}}$ and $A$, as given in Table S1. For each set of structural parameters, at least 7 simulations were initialized from independent fluid or partially crystalline configurations.

To scrutinize phase behavior, we performed simulations in which $A$ and $N_{\textrm{L}}$ were allowed to fluctuate (i.e., sampling an osmotic ensemble \cite{Mehta1994}). These fluctuations were regulated by specified values of pressure $p$ and chemical potential $\mu_{\textrm{L}}$, respectively.  In addition to particle displacement trial moves, osmotic ensemble simulations included trial moves that change $A$ (box area moves) and trial moves that change $N_{\textrm{L}}$ (LHCII insertion and deletion moves). Acceptance probabilities for the latter moves were constructed to satisfy detailed balance, with the area per layer $A$ scaled by the number of layers $n_l$ where necessary.

\subsection{Free energy calculations}
We determined free energy profiles $F(\rho)$ by exploiting the fundamental relationship $F(\rho) = -k_{\rm B}T \ln P(\rho)$ to the probability distribution $P(\rho)$ for spontaneous packing fraction fluctuations. Statistics of $\rho$ were in turn determined by maximum likelihood estimation from a set of 139 osmotic ensemble simulations ($\bar{\mu}_{\textrm{ex,L}} = 0.1$ $k_{\textrm{B}}T$ relative to a standard state; $p = 0.069, 0.070, 0.072$ $k_{\textrm{B}}T/\textrm{nm}^2$) with imposed harmonic bias potentials of the form $u_{\rm bias}(\rho) = k (\rho - \rho_0)^2$. Values of $\rho_0$ and $k$ were chosen to focus sampling on a series of small intervals spanning the range $\rho=0.625$ to $\rho=0.852$ (Fig.~S5). Using the MBAR method as implemented in the PyMBAR package \cite{Shirts2008}, these umbrella sampling results were pooled and reweighted to estimate ${F}(\rho)$ at many pressures and chemical potentials. See SI Methods for details.

\subsection{Clustering algorithm}
PSII arrays were identified using a recursive clustering algorithm with three neighbor criteria, all of which must be satisfied for a PSII particle to be added to a growing array: (i) the angle difference between particle axes must be $<15^{\circ}$; (ii) the component of the center-to-center separation vector parallel to a particle axis must be $<26.5$ nm; (iii) the component of the center-to-center separation vector perpendicular to a particle axis must be $<14$ nm.

\section{Acknowledgments}
We thank K.~Amarnath, A.~Benjamini, N.S.~Ginsberg, M.~Gr\"{u}nwald, H.~Kirchhoff, and P.R.~Shaffer for helpful discussions about the model, and J.D.~Chodera and L.O.~Hedges for advice on free energy calculations. Some computations were performed using resources of the National Energy Research Scientific Computing Center, which is supported by the Office of Science of the U.S. Department of Energy under Contract No. DE-AC02-05CH11231. A.R.S.~was supported in part by a National Science Foundation Graduate Research Fellowship.

\end{document}